\pdfoutput=1

\documentclass[aps,twocolumn,amsmath,amssymb,preprintnumbers,floatfix,prb,superscriptaddress,longbibliography]{revtex4-1}

\usepackage[utf8]{inputenc}
\usepackage{newtxtext}
\usepackage[upint]{newtxmath}
\usepackage{microtype}
\usepackage{textcomp}
\usepackage{eucal}
\usepackage{bm}
\usepackage{siunitx}
\usepackage{comment}
\usepackage{lipsum}

\usepackage{enumerate}
\usepackage{amsfonts}
\usepackage{amsmath}
\usepackage{amssymb}
\usepackage{color}
\usepackage{soul}

\usepackage{graphicx}

\usepackage[colorlinks,allcolors=blue]{hyperref}
\usepackage[capitalize]{cleveref}

\definecolor{DarkRed}{rgb}{0.65,0,0}%
\definecolor{Green}{rgb}{0,0.3,0.3}
\definecolor{Purple}{rgb}{0.3,0,0.65}
\definecolor{Red}{rgb}{1,0,0}
\definecolor{Blue}{rgb}{0,0,0.85}
\definecolor{Magenta}{rgb}{1,0,1}

\newcommand{\ca}[2][]{c_{#2}^{\vphantom{\dagger}#1}} 
\newcommand{\cc}[2][]{c_{#2}^{{\dagger}#1}}          

\newcommand{\be}{\begin{equation}}
\newcommand{\ee}{\end{equation}}

\newcommand{\prlsection}[1]{\textit{#1}.\kern0.05em---\kern0.05em\ignorespaces}

\definecolor{DarkBlue}{rgb}{0,0,0.80}
\definecolor{DarkRed}{rgb}{0.80,0,0}
\definecolor{Purple}{rgb}{0.55,0,0.55}
\definecolor{Purple}{rgb}{0.8,0,0.8}

\begin{document}
\title{Phonon-enhanced optical spin-conductivity and spin-splitter effect in altermagnets}
\author{Erik Wegner Hodt}
\affiliation{Center for Quantum Spintronics, Department of Physics, Norwegian \\ University of Science and Technology, NO-7491 Trondheim, Norway}
\author{Alireza Qaiumzadeh}
\affiliation{Center for Quantum Spintronics, Department of Physics, Norwegian \\ University of Science and Technology, NO-7491 Trondheim, Norway}
\author{Jacob Linder}
\affiliation{Center for Quantum Spintronics, Department of Physics, Norwegian \\ University of Science and Technology, NO-7491 Trondheim, Norway}

\begin{abstract}
Collinear antiferromagnets with nonrelativistic spin-split bands and no net magnetization, called altermagnets, show interesting transport properties due to their unique band structure. We here compute the linear response optical conductivity of thin films of such materials in the presence of phonon scatterings. Using a tight-binding lattice model for altermagnets and the Holstein model for the phonon sector, we find that the electron-phonon scatterings can strongly increase the spin conductivity at finite frequencies. This occurs despite the fact that the self-energy describing the electron-phonon interactions is spin-independent. Interestingly, we show that electron-phonon scattering also enhances the spin-splitter effect at finite frequencies. These results suggest that altermagnets with strong electron-phonon coupling are favorable with regard to AC spin-polarized transport.
\end{abstract}
\maketitle

\section{Introduction}

It has in recent years been understood that collinear antiferromagnets can feature nonrelativistic spin-split electron bands in momentum space despite having compensated magnetic order \cite{Noda-Nakamura-MomentumdependentBandSpin-2016, PhysRevB.99.184432, naka2019spin, Smejkal-Sinova:2020, Hayami-Kusunose-MomentumDependentSpinSplitting-2019, Ahn-Kunes:2019, Yuan-Zunger:2020, Yuan-Zunger-PredictionLowZCollinear-2021} without the requirement of strong electronic correlations \cite{wu_prb_07,  borzi_science_07, classen_prb_20}. Such materials have been classified as a class of antiferromagnets dubbed altermagnets by using a symmetry formalism with operations acting distinctly on spin and real space in crystals \cite{Smejkal-Jungwirth:2022, Smejkal-Jungwirth-ConventionalFerromagnetismAntiferromagnetism-2022, Mazin:2022}. Altermagnets combine properties of ferromagnets and antiferromagnets. While free of a net magnetization and accompanying stray field, altermagnets feature strongly spin-polarized electron bands. This makes them highly interesting for possible applications in spintronics. Altermagnetism is an antiferromagnetic phase with broken combined parity (P) and time-reversal ($\mathcal{T}$) symmetry, leading to non-degenerate spin bands while still preserving the combination of crystal-rotation and time-reversal. It has been theoretically predicted and experimentally verified in several materials \cite{bai2024altermagnetism}.

Spin transport in altermagnets has received much attention since their discovery. This includes not only resistive spin-polarized currents carried by electrons \cite{gonzalez_prl_21, bai_prl_22, karube_prl_22, kolezhuk_prb_24, zarzuela_arxiv_24, hallberg_prb_25}, but also magnonic spin flow \cite{cui_prb_23, hodt_prb_24} and Cooper pair spin flow in superconductors \cite{zyuzin_prb_24, giil_prbl_24, kokkeler_arxiv_24}. However, the role of phonons with regard to spin transport in altermagnets has not been addressed so far. Interplay between different types of quasiparticles, such as phonons and electron-hole excitations, is important to understand the properties of quantum transport in different classes of materials such as altermagnets.

In this work, we compute the frequency-dependent optical conductivity in an altermagnet in the presence of electron-phonon scatterings using linear response Kubo formalism \cite{kubo_jpsj_57}. We use a tight-binding effective model to describe the itinerant electrons in the altermagnet and couple them to optical phonons using a Holstein model \cite{holstein_aop_59}. We consider a 2D system and determine how both the longitudinal spin conductivity and spin-splitter effect are influenced by the phonon scatterings. We find that the electron-phonon coupling can enhance the spin conductivity at finite frequencies. Interestingly, this happens even though the self-energy describing the electron-phonon interaction turns out to be spin-independent within our level of approximation. Moreover, we demonstrate that the spin-splitter effect in altermagnets is affected by the phonon scattering in a similar way. In effect, we find that electron-phonon scattering enhances
the spin-splitter effect at finite frequencies, although the signal at such frequencies is much smaller than in the direct current (DC) limit $\omega =0$. Our findings reveal that substantial electron-phonon coupling in altermagnets could be beneficial with regard to alternating current (AC) spin-polarized transport.

We organize this work as follows. In Sec. \ref{sec:theory}, we introduce the methodology which will be used to obtain the results, namely a Green function approach to the optical conductivity in linear response where the electron-phonon coupling is accounted for via a self-energy. In Sec \ref{sec:results}, we presents results for the AC optical conductivity both for longitudinal spin transport and the transverse spin splitter effect occuring in altermagnets, highlighting the influence of the electron-phonon scattering. Concluding remarks are given in Sec. \ref{sec:conclusion}.

\section{Theory}\label{sec:theory}

We consider a momentum-space square lattice Hamiltonian with a $d$-wave altermagnetic order,
\begin{equation}
H_{\boldsymbol{k}}=h_{\boldsymbol{k}}^{e} + h_{\boldsymbol{k}}^{\alpha}\sigma_z. 
\end{equation}
The 2D free electronic Hamiltonian reads
\begin{equation}
    h_{\boldsymbol{k}}^e = -2t\big[\cos(k_x a) + \cos(k_y a) \big] - \mu
\end{equation}
and the $d$-wave altermagnet interaction is given by 
\begin{equation}
  h_{\boldsymbol{k}}^\alpha = \begin{cases} -\alpha\big[\cos(k_x a) - \cos(k_y a)\big] \quad \text{for $d_{x^2-y^2}$} \\
\hphantom{-}2\alpha\sin(k_x a)\sin(k_y a)\qquad \text{ for $d_{xy}$}
\end{cases}
\end{equation}

\begin{figure}[b!]
    \centering
\includegraphics[width=1.1\linewidth]{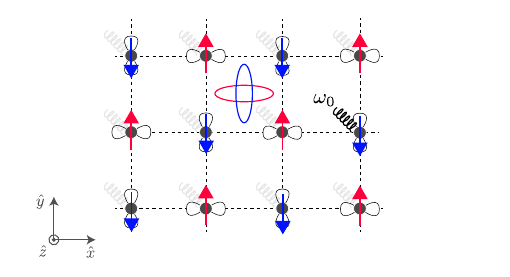}
    \caption{The 2D square lattice hosts an altermagnetic order, modeled by the spin-dependent contribution $h_{\boldsymbol{k}}^\alpha$ arising from a sublattice-dependent crystal environment. The electron density on each site couples locally to a dispersionless Holstein optical phonons with energy $\omega_0$. The Fermi surface shows the $d_{x^2-y^2}$ variant of the altermagnetic order, and by driving a charge current along the lattice diagonals, a transverse spin-polarized current will arise due to the spin-splitter effect. }
    \label{fig:lattice}
\end{figure}
\begin{figure}
    \centering
\includegraphics[width=0.9\linewidth]{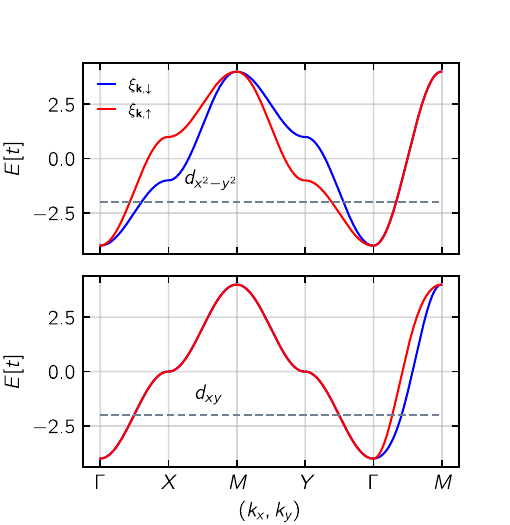}
    \caption{Electronic bandstructure for a $d_{x^2-y^2}$ altermagnet (upper panel) and a $d_{xy}$ (lower panel) for $\alpha=0.25t$. The grey stippled line denotes the chemical potential $\mu=-2$ used in the optical conductivity calculations.}
\label{fig:bandstructure}
\end{figure}

depending on whether we want to describe $d_{x^2-y^2}$ or $d_{xy}$ altermagnetic order, the latter obtained by rotating the lobes of the Fermi surface in the $d_{x^2-y^2}$ case (shown in Fig. \ref{fig:lattice}) in momentum space by $\pi/4$. The electronic band structure for these two types of order is shown in Fig. \ref{fig:bandstructure}. The quantity $\alpha$ parametrizes the altermagnetic interaction between the itinerant electrons and the localized spins. From now on, we measure lengths in $a$ and energies in $t$, thus effectively setting $a=t=1$ and also using natural units $\hbar=c=1$. The Néel vector direction has explicitly been set to $\hat{\boldsymbol{z}}$ such that the Hamiltonian is spin-diagonal and the dispersion is given by 
\begin{equation}
    \xi_{\boldsymbol{k},s}=h_{\boldsymbol{k}}^e + s {h}_{\boldsymbol{k}}^{\alpha}
\end{equation}
for the two spin sub-bands labeled by $s$.

\subsection{Electron-phonon self-energy}

The electrons are assumed to be coupled to Holstein dispersionless optical phonons of energy $\omega_0$ through the coupling Hamiltonian
\begin{equation}
    H_{\text{e-p}}=\frac{\gamma}{\sqrt{N}}\sum_{\boldsymbol{k},\boldsymbol{q},\sigma}\cc{\boldsymbol{k}+\boldsymbol{q},\sigma}\ca{\boldsymbol{k},\sigma}(a_{\boldsymbol{q}}^{\vphantom{\dagger}}+a_{-\boldsymbol{q}}^\dagger)
\end{equation}
where $N$ is the number of unit cells in our 2D system and where $\gamma$ is a coupling constant of unit energy into which we have subsumed the ionic mass $M$ and the phonon energy $\omega_0$. Here, the fermionic $c$-operators govern the electronic degrees of freedom with momentum $\boldsymbol{k}$ and spin projection $\sigma$ while the bosonic $a$-operators create and annihilate magnons with wave vector $\boldsymbol{q}$. Treating the coupling as a perturbation, the contribution to the self energy $\Sigma(\omega)$ of lowest order in $\gamma$ is given by the $2\times 2$ matrix
\begin{multline}
    \Sigma(i\omega_l)=\\-\frac{1}{\beta}\frac{\gamma^2}{N}\sum_{\boldsymbol{q}, i\nu_l}D^{(0)}(\boldsymbol{q}, i\nu_l)G^{(0)}(\boldsymbol{k}-\boldsymbol{q}, i\omega_l - i\nu_l) \label{eqn:selfenergy}
\end{multline}
The bare phonon propagator reads
\begin{equation}
    D^{(0)}(\boldsymbol{q},i\nu_l)=\frac{2\omega_0}{(i\nu_l)^2 - \omega_0^2}
\end{equation}
where $\nu_l$ is a bosonic Matsubara frequency. The propagator is thus momentum independent in the Holstein model, such that the self-energy in Eq. (\ref{eqn:selfenergy}) becomes momentum-independent as well. The momentum sum runs over the first Brillouin zone of the square lattice. The $2\times2$ bare electron propagator is diagonal in the spin basis and the spin-dependent Green's functions are given simply by
\begin{equation}
    G_{s}^{(0)}(\boldsymbol{k}, i\omega_l)=\frac{1}{i\omega_l - \xi_{\boldsymbol{k}, s}} \label{eqn:bareG}
\end{equation}

 Upon insertion of Eq. (\ref{eqn:bareG}) into Eq. (\ref{eqn:selfenergy}), we obtain 
 \begin{equation}
     \Sigma(i\omega_l)=\begin{bmatrix}\Sigma_{\uparrow}(i\omega_l) & 0 \\ 0 & \Sigma_{\downarrow}(i\omega_l)  \end{bmatrix}
 \end{equation}
 where the spin-dependent self-energy reads
 \begin{multline}
     \Sigma_s(i\omega_l)=\frac{\gamma^2}{N}\sum_{\boldsymbol{q}}\bigg[\frac{N_0 +n_\text{F}(\xi_s)}{i\omega_l -\omega_0 -\xi_s}\\ + \frac{N_0 + 1 - n_{\text{F}}(\xi_s)}{i\omega_l + \omega_0 - \xi_s} \bigg] \label{eqn:selfenergy}
 \end{multline} 
and $N_0=n_\text{B}(\omega_0)$ is the Bose-Eisenstein distribution. While the self-energy appears to be spin-dependent at first glance, the self-energy is effectively spin-independent due to the compensated nature of the altermagnetic order. This can be seen by a simple rotation of the momentum domain $\boldsymbol{q}$ in the summand of Eq. (\ref{eqn:selfenergy}) by $\pi/2$ which effectively maps $\Sigma_{\uparrow}\rightarrow\Sigma_{\downarrow}$. From now on, we will therefore omit the spin index on the self-energy. 

The total interacting Green's function for a given momentum mode is given by
 \begin{align}
     G^{-1}&=(G^{(0)})^{-1}-\Sigma(i\omega_l)\mathbb{I}_2 \\
     \Rightarrow G&=\big[i\omega_l\mathbb{I} - H_{\boldsymbol{k}} - \Sigma(i\omega_l)\mathbb{I}_2 \big]^{-1} 
 \end{align}
 where the real and imaginary parts of the self-energy provide a correction to the energy and lifetime of the quasiparticles. The total Green's function is thus diagonal in spin space, and its diagonal elements can be written as 
 \begin{equation}
     G_s(i\omega_l, \boldsymbol{k})=\frac{1}{i\omega_l - \xi_s - \Sigma(i\omega_l)}
 \end{equation}

\subsection{Optical conductivity in  altermagnets}
We now obtain an expression for the spin-resolved optical conductivity $\sigma_{nm}^s(\omega)$ which quantifies the current response with spin polarization \textit{s} in direction $n$ due to an applied electric field in direction $m$, where $n,m\in\{x,y\}$. In the absence of spin-orbit interactions, the spin-polarized current is conserved in the continuity equation when using the standard definition of a spin current with a velocity operator that is modified by the altermagnetic term. In this case, spin-polarized transport can be computed from the spin-resolved optical conductivity $\sigma_s(\omega)$ for charge simply as the difference between charge conductivity for spin-$\uparrow$ and spin-$\downarrow$ carriers. We consider this scenario below. In a linear response approach, the optical conductivity tensor is given by, 
\begin{equation}
    \sigma_{nm}(\omega)=\frac{i}{\omega}\big[\Pi_{nm}(\omega)-\Pi_{nm}(0)\big]
\end{equation}
where we have confined our inquiry to the homogeneous response ($\boldsymbol{q}=0$). The Matsubara current-current correlation function is given by
\begin{widetext}
\begin{equation}
    \Pi_{nm}(i\omega_l, \boldsymbol{q}=0)=\frac{1}{\beta}\int\frac{d^2 k}{(2\pi)^2}\sum_{\nu_l}\text{Tr}\bigg[\hat{j}_{n}(\boldsymbol{k})G(i\nu_l, \boldsymbol{k})\hat{j}_{m}(\boldsymbol{k})G(i\nu_l+i\omega_l, \boldsymbol{k}) \bigg] \label{eqn: correlation function}
\end{equation}
\end{widetext}
where the current operator $\hat{j}_n(\boldsymbol{k})$ is given as 
\begin{equation}
\hat{j}_n(\boldsymbol{k})=-\frac{e}{\hbar}\frac{\partial }{\partial k_n}H_{\boldsymbol{k}}=-\frac{e}{\hbar}\frac{\partial h_{\boldsymbol{k}}^e}{\partial k_n}\mathbb{I}_2 -\frac{e}{\hbar} \frac{\partial {h}_{\boldsymbol{k}}^{\alpha}}{\partial k_\alpha}\sigma_z\label{eqn: current operator}
\end{equation}
The current operator remains unaffected by the presence of phonons, something which can be understood from the fact that the matrix element in the Hamiltonian term describing such scattering is momentum-independent in the Holstein model. Such a term does not influence the definition of a current operator, since it commutes with the number operator $c_{\boldsymbol{k}}^\dag c_{\boldsymbol{k}}^{\vphantom{\dag}}$. 
The spin-diagonal nature of both our Green's functions and current operator causes the correlation function to be easily evaluated by the insertion of $G$ and $\hat{j}$,
\begin{widetext}
    \begin{equation}
        \Pi_{nm}^\text{intra}(i\omega_l)=\frac{e^2}{\hbar^2\beta}\int \frac{d^2 k}{(2\pi)^2}\sum_{\nu_l}\sum_{s}\frac{\partial \xi_{\boldsymbol{k},s}}{\partial k_n}\frac{\partial\xi_{\boldsymbol{k}, s}}{\partial k_m}G_s(i\nu_l, \boldsymbol{k})G_s(i\nu_l + i\omega_l, \boldsymbol{k})
    \end{equation}
\end{widetext}
The absence of scattering between spin states causes the correlation function to only contain an intraband contribution, which we have specified through the superscript on $\Pi_{nm}$ above.

The presence of the self energy $\Sigma(i\nu_l)$ makes a direct evaluation of the Matsubara sum difficult. We tackle this by rewriting the sum as
\begin{widetext}
    \begin{align}
    \frac{1}{\beta}\sum_{\nu_l}G_s(i\nu_l, \boldsymbol{k})G_{{s}}(i\nu_l + i\omega_l, \boldsymbol{k})&=\int \frac{d\omega_1}{2\pi}\frac{d\omega_2}{2\pi}A_{s}(\omega_1, \boldsymbol{k})A_{{s}}(\omega_2, \boldsymbol{k})\frac{1}{\beta}\sum_{\nu_l}\frac{1}{\omega_1 - i\nu_l}\frac{1}{\omega_2 - i\nu_l - i \omega_l} \\
    &=  \int \frac{d\omega_1}{2\pi}\frac{d\omega_2}{2\pi}A_{s}(\omega_1, \boldsymbol{k})A_{{s}}(\omega_2, \boldsymbol{k})\frac{n_\text{F}(\omega_1) - n_\text{F}(\omega_2)}{i\omega_l + \omega_1 - \omega_2}
\end{align}
\end{widetext}
where we used the relation
\begin{align}
    G(\boldsymbol{k}, i\omega_l)=\int \frac{d\omega}{2\pi}\frac{A(\boldsymbol{k}, \omega)}{i\omega_l - \omega}.
\end{align}
Taking the $T\rightarrow0$ limit and performing the analytical continuation of the Matsubara frequency, the current-current correlation function then become 
\begin{widetext}
    \begin{equation}
        \Pi_{\alpha\beta}^\text{intra}(\omega)=\frac{e^2}{\hbar^2}\int \frac{d^2 k}{(2\pi)^2}\sum_{s}\frac{\partial \xi_{\boldsymbol{k},s}}{\partial k_\alpha}\frac{\partial\xi_{\boldsymbol{k}, s}}{\partial k_\beta}\int \frac{d\omega_1}{2\pi}\frac{d\omega_2}{2\pi}A_s(\omega_1, \boldsymbol{k})A_{{s}}(\omega_2, \boldsymbol{k})\frac{\theta(\omega_2) - \theta(\omega_1)}{\omega + \omega_1 - \omega_2 + i\eta} \label{eqn:intra}
    \end{equation}
\end{widetext}
where the spectral weight is given by
\begin{align}
    A_s(\omega, \boldsymbol{k})&=-\frac{1}{\pi}\text{Im}\{G_s^R(\omega, \boldsymbol{k})\} \\
    &=-\frac{1}{\pi}\frac{\eta + \text{Im}\{\Sigma(\omega) \}}{(\omega - \xi_{\boldsymbol{k}, s} - \text{Re}\{\Sigma(\omega) \})^2 + (\eta + \text{Im}\{\Sigma(\omega) \})^2}
\end{align}
where $\eta\to0^+$ is an infinitesimal, real positive quantity arising from the analytical continuation of the Matsubara Green function while $\theta(\omega)$ is a step-function.

\section{Results}\label{sec:results}
For a general response function connecting a physical observable and applied stimulus, it is the relative time-reversal parity of the observable and stimulus 
that determines whether the real and imaginary parts of the response function are reactive or dissipative. Consider the longitudinal current response to an applied dc electric field. For a time-reversal odd charge current $J$ satisfying Ohm's law, the conductivity $\sigma$ must be time-reversal odd since the dc electric field $E$ is time-reversal even. The relative time-reversal parity of observable and stimulus is opposite, and the real part of the conductivity describes a dissipative process. 

\begin{figure}
    \centering
    \includegraphics[width=0.9\linewidth]{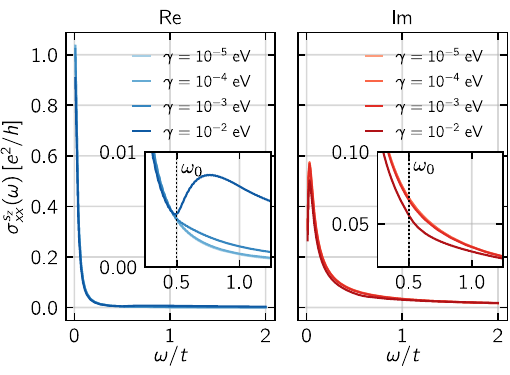}
    \caption{Longitudinal AC spin conductivity $\sigma_{xx}^{S_z} = \sigma_{xx}^\uparrow - \sigma_{xx}^\downarrow$ for a $d_{x^2-y^2}$ altermagnet with $\alpha=0.25t$ for several strengths of the electron-phonon coupling. Due to the $\mathcal{T}\mathcal{C}_4$ relation between the spin-up and -down Fermi surfaces of the altermagnet, it follows that $\sigma_{yy}^{S_z}=-\sigma_{xx}^{S_z}$. Here, $\mathcal{C}_4$ represents the operation of a lattice rotation by $\pi/4$.}
    \label{fig:longspin}
\end{figure}

For a time-dependent stimulus (AC electric field), there can be both an in- and out-of-phase contribution to the observable (charge current). This is conveniently described using a complex conductivity. The real component of the complex conductivity still determines the dissipation in the system, whereas the imaginary component determines the dissipationless (reactive) part of the response. This is shown by a direct computation of the rate of energy change $\langle \dot{H}\rangle$ in the system. Invoking time-reversal symmetry arguments to conclude with this is less clear than in the dc case for an intrinsically time-dependent electric field, since the explicit time-dependence of $E$ can make the electric field odd under time reversal. In what follows, we consider both the dissipative and reactive parts of the AC conductivity in an altermagnet in the presence of electron-phonon scattering. In the absence of spin-orbit coupling, the electrical conductivity is independent of the orientation of the Néel order parameter. We will show that the electron-phonon scattering can strongly increase the longitudinal spin
conductivity at finite frequencies. Secondly, we consider how the spin-splitter effect in an altermagnet is affected by electron-phonon scattering. In particular, we find that the spin splitter effect inherits the strong enhancement of the underlying longitudinal spin conductivities indicating that electron-phonon coupling can significantly enhance the AC spin splitter response in altermagnets. 
\begin{figure}
    \centering
\includegraphics[width=0.9\linewidth]{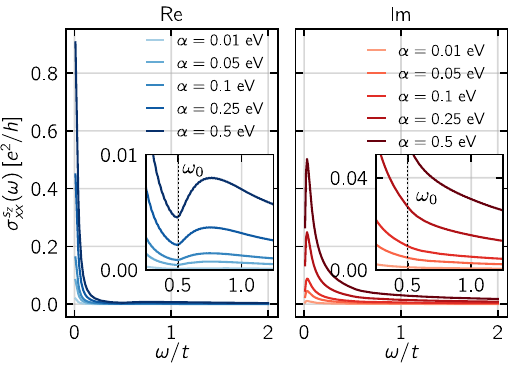}
    \caption{Longitudinal AC spin conductivity $\sigma_{xx}^{S_z} = \sigma_{xx}^\uparrow - \sigma_{xx}^\downarrow$ for a $d_{x^2-y^2}$ altermagnet with fixed electron-phonon strength $\gamma=0.01$ eV for several strengths of the altermagnetic order.}
    \label{fig:longam}
\end{figure}


\begin{figure}
    \centering
    \includegraphics[width=0.9\linewidth]{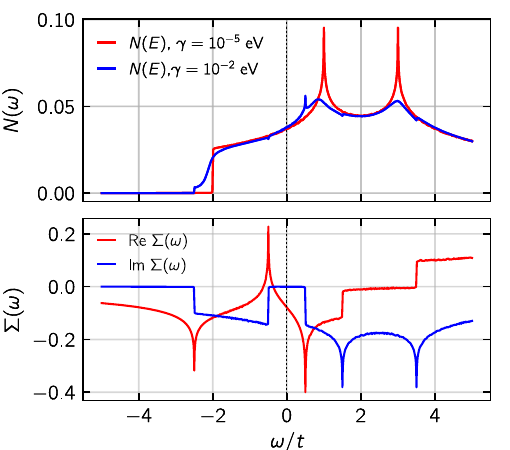}
    \caption{Upper panel: Density of states $N(\omega)$ computed from the spectral function with a small electron-phonon interaction $\gamma = 10^{-5}$ eV (red line) and a large interaction $\gamma = 10^{-2}$ eV (blue line). Lower panel: Real and imaginary part of the phonon self-energy. A $d_{x^2-y^2}$ dispersion with $\alpha=0.25t$ was used while the dotted vertical line denotes the chemical potential. All energies are measured from the chemical potential. }
    \label{fig:DOSandSigma}
\end{figure}

\subsection{AC longitudinal spin conductivity}

The spin-dependent Fermi surface of the altermagnet gives rise to a spin-dependent group velocity and a subsequent spin-polarized charge current. As long as the direction of the applied field does not align perfectly with one of the nodal directions of the altermagnetic Brillouin zone, this property will manifest as a nonzero longitudinal spin conductivity,
\begin{equation}
    \sigma_{xx}^{S_z}=\sigma_{xx}^\uparrow - \sigma_{xx}^\downarrow
\end{equation}
and likewise for the \textit{y}-direction where $\sigma_{xx}^{\uparrow/\downarrow}$ are the spin-dependent charge conductivities for $\uparrow/\downarrow$ spins.

In Fig. \ref{fig:longspin}, we show the optical spin conductivity for a $k_x^2-k_y^2$ altermagnet for an altermagnet strength $\alpha=0.25t$. No spin-orbit coupling is present, and the only contribution to the longitudinal spin conductivity arises from intraband processes described by Eq. (\ref{eqn:intra}). The introduction of electron-phonon coupling significantly modifies the spin conductivity for frequencies above the phonon frequency $\omega_0$ as phonon-mediated processes become accessible. This gives rise to a so-called phonon sideband \cite{PhysRevB.81.045419} in the AC conductivity due to the increased absorption. Physically, this may be understood as follows. In the absence of interband scattering, the optical conductivity stems solely from horizontal intraband transitions where a photon kicks an electron in a filled state up to an available state. While this happens with momentum conservation, as seen from Eq. (\ref{eqn:intra}), the finite broadening provided by the imaginary part of the phonon self-energy smears the electron band due to the broadening of the spectral weight $A(\omega,\boldsymbol{k})$ and thus enables momentum-conserving energy transitions even for a single band. The electron density of states $N(\omega)$ and phonon self-energy $\Sigma(\omega)$ are shown in Fig. \ref{fig:DOSandSigma}. The imaginary part of $\Sigma$ is absent for $|\omega|<\omega_0$, signifying that the electron-phonon scattering requires a minimum energy of $\omega_0$ before it can provide a decay channel and thus a finite lifetime for an electron state. The real part of the self-energy remains finite even for $\omega<\omega_0$ and causes a renormalization of properties such as Fermi velocity and chemical potential of the electrons. The distinct features of the self-energy at $\omega_0$ are manifested also in the DOS $N(\omega)$ in the form of a peak at the same frequency and lead to an enhancement of the optical conductivity near $\omega=\omega_0$.  

For frequencies $|\omega|<\omega_0$, the only broadening present is due to the $i\eta$ introduced to ensure causality of the Green's functions, being in our calculations of a large magnitude $10^{-2}t$ as a simple model for other sources of broadening due to scattering besides electron-phonon interactions. In the limit of a clean metal, $\eta\rightarrow0^+$ and the optical conductivity should, in the absence of altermagnetic order, approach the Drude peak $\sigma(\omega)\propto \delta(\omega)$. In effect, without the finite lifetime induced by electron-phonon coupling, intraband scattering processes that conserve momentum vanish as $\eta \to 0^+$, since the spectral weight becomes sharply defined and its smearing disappears.

In Fig. \ref{fig:longam}, we also show the longitudinal spin conductivity for a fixed electron-phonon strength $\gamma=0.01$ eV as a function of the altermagnetic order strength. As expected, the spin conductivity vanishes with vanishing altermagnet strength. The values used for the electron-phonon coupling strength in all figures are sufficiently small $(\gamma/t \ll 1)$ that they lie in the regime where vertex corrections can be neglected \cite{cichutek_prb_22}.

\subsection{AC Spin-splitter effect}
The spin-splitter effect refers to the phenomenon of a charge current inducing a transverse spin-polarized current whenever the direction of the applied charge current does not align with the directions of the altermagnetic lobes. In terms of the spin-resolved conductivity, the spin-splitter effect can be expressed as $    \sigma_{xy}^\uparrow = -\sigma_{xy}^\downarrow$, allowing for the definition of a spin-splitter conductivity
\begin{align}
\sigma_{xy}^{S_z} = \sigma_{xy}^\uparrow - \sigma_{xy}^\downarrow.
\end{align}

\begin{figure}
    \centering
    \includegraphics[width=0.9\linewidth]{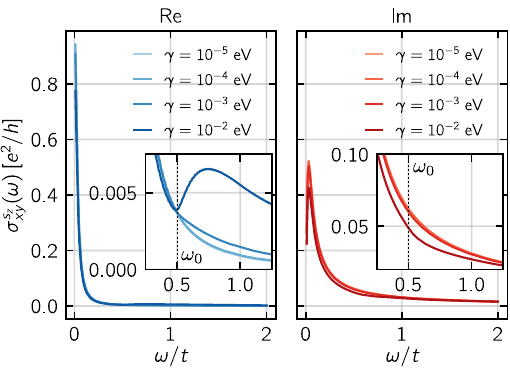}
    \caption{AC spin-splitter effect $\sigma_{xy}^{S_z} = \sigma_{xy}^\uparrow - \sigma_{xy}^\downarrow$ for a $d_{xy}$ altermagnet with $\alpha=0.25t$ for several strengths of the electron-phonon coupling.}
    \label{fig:spinsplitter}
\end{figure}

The emergence of the spin-splitter effect \cite{gonzalez_prl_21, Bai2022, Karube2022, zarzuela_prb_25} can be attributed to the anisotropic spin-dependent conductivities. Due to the $\mathcal{T}\mathcal{C}_4$ relation between the spin-up and -down \textit{d}-wave Fermi surface, a particular spin species experiences an anisotropic conductivity reflected through $\sigma_{xx}^\sigma(\omega) \neq \sigma_{yy}^\sigma(\omega)$. Here, $\mathcal{C}_4$ represents the operation of a lattice rotation by $\pi/4$. When applying an electric field along a direction away from the directions along which the altermagnetic lobes are elongated, a transverse current response arises, known as the spin-splitter effect. 
With a $d_{xy}$ altermagnet, we obtain a non-zero $\sigma_{xy}^{S_z}$, depicted as a function of electron-phonon strengths in Fig. \ref{fig:spinsplitter}. The spin-splitter effect essentially emerges from the underlying anisotropies of the spin-dependent charge conductivities and, as such, the transverse spin conductivity inherits the electron-phonon induced enhancement in the dissipative part for frequencies $\omega>\omega_0$. It should nevertheless be remarked that the magnitude of the spin-splitter effect at the frequencies where it is enhanced by the phonon scattering is considerably smaller than in the DC limit. 

\section{Concluding remarks}\label{sec:conclusion}
We have calculated the frequency-dependent
conductivity in an altermagnet with electron-phonon scattering
using the Kubo formula and the Holstein phonon model. Both longitudinal spin conductivity and the spin-splitter effect are enhanced substantially at finite frequencies. 
This indicates that electron-phonon coupling in altermagnets could be useful to enhance AC spin currents.

\acknowledgments

We thank N. H. Aase for useful discussions. This work was supported by the Research
Council of Norway through Grant No. 323766 and its Centres
of Excellence funding scheme Grant No. 262633 “QuSpin.” Support from
Sigma2 - the National Infrastructure for High Performance
Computing and Data Storage in Norway, project NN9577K, is acknowledged.

\bibliography{refs.bib}

\appendix

\end{document}